# Towards Adaptable and Adaptive Policy-Free Middleware

Alan Dearle, Graham N.C. Kirby, Stuart J. Norcross, Angus D. Macdonald and Greg J. Bigwood
School of Computer Science, University of St Andrews
North Haugh, St Andrews, Fife KY16 9SX, Scotland
{al, graham, stuart}@cs.st-andrews.ac.uk, {adm15, gjb4}@st-andrews.ac.uk

## ABSTRACT
We believe that to fully support adaptive distributed applications, middleware must itself be *adaptable*, *adaptive* and *policy-free*. In this paper we present a new language-independent adaptable and adaptive policy framework suitable for integration in a wide variety of middleware systems. This framework facilitates the construction of adaptive distributed applications. The framework addresses adaptability through its ability to represent a wide range of specific middleware policies. Adaptiveness is supported by a rich contextual model, through which an application programmer may control precisely how policies should be selected for any particular interaction with the middleware. A contextual pattern mechanism facilitates the succinct expression of both coarse- and fine-grain policy contexts. Policies may be specified and altered dynamically, and may themselves take account of dynamic conditions. The framework contains no hard-wired policies; instead, all policies can be configured.

## 1. INTRODUCTION
Middleware is becoming increasingly important in today's information society, providing the glue holding together the many distributed systems and services on which we rely. To quote the call for this conference, we believe that middleware infrastructure is becoming "more and more heterogeneous and complex". There is a belief by many that we already have enough middleware, and that middleware is a solved problem. However, current middleware is limited in its ability to support adaptive applications, for reasons that will be explained below. As identified in the call, adaptiveness cannot simply be added to a system like a plug-in module. In our approach, the ability to support adaptive applications forms a core part of the middleware architecture. In this paper we focus on the wide class of middleware systems based on a synchronous remote invocation model, including RPC, RMI and SOA style middleware.

We observe three trends in such systems: divergence, increasing complexity, and lack of flexibility. Recently remote invocation style middleware systems have separated into two broad areas—systems based on the distributed object model (DOM), and service-oriented systems. It has been argued [1, 20] that service-oriented computing is fundamentally different from distributed object computing. Nonetheless, there is no intrinsic requirement for middleware to take a position on which model is adopted for a particular application, or indeed to restrict the application to one model. Instead, the middleware should be capable of adapting its behaviour, dynamically if necessary, to the application requirements. For example, middleware should be adaptable enough to allow a remotely accessible component to support interaction with both DOM-style and service-oriented clients (such as Web Services clients) as and when appropriate.

The trend towards inflexibility is largely an historical accident. Most common middleware systems embody fixed policies governing the interaction between components in a distributed application. The application programmer must take account of these policies at system design time. For example, when distributed applications are written using object-oriented middleware such as DCOM [3], .Net [14], CORBA [16] or Java RMI [6] the programmer decides how the application is to be distributed across multiple machines and (statically) identifies various categories of application classes. These categories include those classes that have the ability to be accessed remotely; those that may be transmitted over the network; and those that are unable to take part in inter-address-space interactions.

The consequence of the above trends is that applications written using middleware are brittle and it is difficult to achieve adaptiveness. There is no way, for example, for data to be passed *by value* when devices are attached via a high bandwidth connection and *by reference* when poorly connected. Similarly, a space-limited device cannot request data *by reference* while a less constrained workstation requests data *by value*. These simple examples demonstrate the motivation for our work.

We believe that to fully support adaptive distributed applications, middleware must itself be *adaptable*, *adaptive* and *policy-free*. Adaptable middleware is capable of being configured to improve its suitability for a particular situation or environment. This is essential if applications are to be able to adapt their interaction patterns to circumstances, since they rely on the middleware to perform all remote interaction. Adaptive middleware can configure itself automatically. This is needed in order to support fine-grain adaptation to dynamic circumstances, since by definition this cannot be addressed by static configuration policies. Policy-free middleware contains no hard-wired policies; instead, all policies can be configured. This is essential if applications are to be permitted to adapt over the full spectrum of the policy space.

In this paper we present a new language-independent adaptable and adaptive policy framework suitable for integration in a wide variety of middleware systems, supporting the construction of adaptive distributed applications. The framework addresses adaptability through its ability to represent a wide range of specific middleware policies, of which data transmission *by value* and *by reference* are simple examples.

Adaptiveness in middleware may be characterised as the ability of the middleware to make autonomous decisions based on context. Such decisions should not be based on fixed rules; instead, the programmer should be able to precisely control how policies should be selected in any particular circumstance. In the framework described in this paper, these circumstances are described using a rich contextual model. A contextual pattern mechanism facilitates the succinct expression of both coarse- and fine-grain policy contexts. Policies may be specified and altered dynamically, and may themselves take account of dynamic conditions.

We have prototyped this framework in the context of the Java-based RAFDA middleware system [5]. The examples shown in this paper are in Java, but the concepts supported by the framework are applicable to any language.

## 2. MIDDLEWARE DESIGN ISSUES

### 2.1 Design Principles

In the introduction we advocated the need for flexible and adaptive middleware. Such middleware should be compliant [15] with the application requirements rather than the application being coded in a manner that is compliant with the middleware. The manner in which inter-address-space interactions occur should be controllable via the specification of policies that may be separated from the code. Many of the motivations for this are similar to the arguments for Aspect-Oriented Programming (AOP) [11]. Firstly, it is good software engineering practice to separate concerns. Secondly, we wish to be able to reuse pre-existing code (including library code), which precludes changing code in order to make calls into the middleware. This also requires that the middleware should impose no restrictions on application structure or the type hierarchy of application classes. Lastly, as argued above, the middleware should be compliant with the application, permitting application designers to fine-tune middleware behaviour. Some examples of adaptability are inherently data-centric. For example, in some application, it may be desirable to pass large objects and small ones using different encoding mechanisms.

Our aim is therefore to create a middleware framework that is adaptive, simple, efficient, extensible and programmable. We have already discussed the need for adaptiveness. Simplicity requires that common policy decisions can be made with little or no programmer involvement. For example it should be trivial to configure the framework to replicate the policies of conventional distributed object and service-oriented models. The need for efficiency is obvious; highly adaptive but slow middleware systems are unlikely to have any uptake—middleware mechanisms therefore need to be both tractable and efficient. Extensibility and programmability require that arbitrarily complex interactions may be specified, including the injection of application programmer code into the middleware system.

These design principles could be applied to various middleware systems in a number of ways. In our approach we structure the adaptive middleware design space as follows:

- **flexibility dimensions**: those aspects of middleware behaviour where commonly-imposed restrictions are undesirable;
- **policy dimensions**: those aspects of middleware behaviour for which adaptation is desirable;
- **meta-policy dimensions**: those aspects of execution context that may be used to determine policy selection.

The following sections contain a non-exhaustive list of examples of each of these sets of dimensions.

### 2.2 Flexibility Dimensions

The dimensions of flexibility represent adaptiveness requirements over various aspects of middleware behaviour. Restrictions imposed by common middleware systems in these areas limit the ability of applications to exhibit adaptiveness.

One example of such a dimension is that of application distribution boundaries. To accommodate adaptation, objects of arbitrary classes should be able to participate in inter-address-space calls, and be able to be transmitted across the network. This contrasts with most middleware systems, as outlined in Section 1.

Another dimension is in the rules governing type equivalence. Most middleware systems use name equivalence when matching the types of local and remote objects[1]. Furthermore, those names are bound to statically defined interfaces and classes. Consider the example shown in Figure 1:

```
class Student extends Person
  implements Scholar {...}
```

**Figure 1. Class Student.**

In many middleware systems a remote instance of *Student* can be accessed remotely as a *Student*, *Person* or *Scholar*[2]. However, if some application needs to export instances typed as *Human*, which is structurally compliant with the *Student* class, it cannot, since *Student* was not defined to extend *Human*. Clearly, this limits adaptiveness since the static definition is often outwith the control of application programmer and cannot be changed once the class has been defined. One solution to this problem is to use structural type matching [4]. Using structural matching, an object may be remotely accessed using any type that is compatible with the structure of the implementing type.

In general, there are three types with which the programmer is concerned: the concrete type of the remotely accessible object; the type with which it is made accessible; and the type by which the remote object is viewed. Any restriction beyond structural compatibility between these types is undesirable. In the previous example, if the application programmer has defined a type *Human* after the fact, it should be possible for the *Student* instance to be made remotely accessible using this interface, even though *Student* does not explicitly extend or implement *Human*.

A third example of a dimension of flexibility is in the sub-typing rules applying to object transmission. In most middleware systems, the manner in which values are encoded is dictated by the static types defined in the remote interface. For example, using standard Web Services, each service method has a statically defined return type and can return only objects that are of exactly this type. The reasons for this restriction are complex [21]. However, for maximum flexibility, arbitrary sub-typing should be accommodated, allowing sub-types to be returned by interface methods where appropriate.

### 2.3 Policy Dimensions

The dimensions of policy represent the aspects of middleware behaviour that can be configured. Again, most middleware systems impose significant restrictions on these, limiting application adaptiveness. Possible dimensions include:

---

[1] Some systems are more restrictive than this, e.g. CORBA requires the same IDL to be used on the client and server.

[2] In some systems these classes must extend special interfaces or classes such as *RemoteObject* or *MarshalByRefObject*.

- object transmission
- object encoding
- object placement
- security

The first and most obvious policy dimension is *transmission policy*: how objects are transmitted between address-spaces. Possible policies include *by value*, *by reference*, *by move* and *by visit*. In most middleware systems the transmission policy is fixed. For example, using Web Services transmission of objects is always *by value* whereas in Java RMI the transmission policy is determined by the type and interfaces of the object being transmitted.

A programmer may also wish to apply a specific transmission policy to part or all of the closure of an object being transmitted. For example, some object closure might be passed *by value* to a given depth, and *by reference* thereafter. Such policies have two potential benefits—preventing large object closures from being transmitted *by value*, and obviating the need to copy data from one data structure to another when it is exported via a service. Hybrid transmission policies are also desirable, in which objects are passed by reference but some fields of those objects are passed by value along with the reference and cached by clients.

Closely related to transmission policy is *encoding policy*, which determines the manner in which values are encoded when they are sent between address spaces. For example, an object transmitted *by value* may be encoded using SOAP or as a CORBA IIOP message. Even where a single base encoding mechanism such as SOAP is used, it may be advantageous to vary the encoding of values in certain circumstances. For example, a large array might be *base64* encoded within a SOAP message when transmitted to a peer which could cope with such an encoding, with a standard encoding being used otherwise.

The next dimension is *placement policy*, which controls where objects are placed. There are two related aspects: *creation policy* and *migration policy*. The first of these controls in which address space objects are created, the second where, and under which circumstances, they are migrated. Policies governing these aspects are restrictive in most current middleware systems—typically all objects are created in the same address space as their creator, and do not migrate between address spaces. Exceptions discussed in Section 5 include [8, 13, 17, 19]; whilst these are flexible with respect to placement policy, they do not address all of the policy dimensions discussed here.

The dimension of *security policy* encompasses various mechanisms including policies used to control access to objects from remote address-spaces and link-level security between hosts. As with the other dimensions, for truly adaptive applications it may be necessary to vary security policies on a fine-grain basis.

## 2.4 Meta-Policy Dimensions

Meta-policies determine the circumstances in which particular policies should be applied. The dimensions comprise the various aspects of the execution context that might be relevant in choosing the appropriate policy in a given situation. The context varies between policy dimensions: for transmission and encoding policy it is the context of object transmission; for placement policy it is the context of object instantiation, and so on.

For brevity we focus here on meta-policy dimensions for transmission and encoding policies. Possible dimensions include:

- **the execution environments of the client and server**[3]: for example, the client might be a particular version of a web browser, or the server might be a web server.
- **the identities of the client and server**: for example, an application might interact with one server *by value* because it is a conventional Web Services server, and with another using a hybrid caching policy which that server supports.
- **the service being accessed**: two services provided in the same address-space might be configured to behave differently. Conversely, an application calling services on two different machines might wish to interact with each differently in accordance with load, latency, size of data etc.
- **the class of the object being transmitted**: this might be used to transmit *by value* all instances of a class known to have immutable fields.
- **the identity of the object being transmitted**: this permits policy to be applied to individual object instances. This might be used to treat large objects differently from small ones, or to support application-specific caching policies.
- **the name of the parameter/return value being transmitted**: this allows fine-grain policy to be applied to individual method parameters and to the return value.
- **the method being executed**: this allows the same policy to be applied to all parameters and the return value of a method.
- **the current thread**: this allows policy to be applied only to a particular thread. This might be used when different server threads are being used to service two different applications running within a single address-space.
- **the package of the class of the object being transmitted**: the same policy can be applied to all classes in a package.
- for objects being transmitted within the closure of another:
    - **the type** of the original object.
    - **the name of the field** in the referring object through which the current object was reached.

These dimensions are clearly not mutually independent—for example, the *class* and *package* dimensions. Rather than aiming for absolute minimality, we consider it preferable to support a rich set of dimensions, in order to give the programmer flexibility. Given the large size of this set, we need mechanisms to enable both coarse- and fine-grain meta-policies to be defined succinctly. Key to this is the ability to easily specify sensible default meta-policies to be adopted in the absence of more specific ones. As previously noted, however, these defaults should not be hard-wired into the middleware, which should itself be policy-free.

## 2.5 Policy Framework Requirements

Given the approach outlined in this section, the main requirements of an adaptive policy framework may be summarised as follows:

- to allow programmers to select pre-defined policies and to specify new ones, within each of the policy dimensions;
- to allow programmers to define new meta-policies, each specifying a context in terms of one or more of the relevant meta-policy dimensions, and a corresponding policy;
- whenever the middleware system is about to perform an action governed by policy: to automatically identify the meta-policy

[3] Our framework may be used in P2P environments. To aid readability we refer to the caller as the *client* and the callee the *server*.

whose context specification most closely matches the current context, and to enact the corresponding policy.

Our approach is described in Section 4.

# 3. USE CASES

This section gives motivating examples for an adaptive policy-free middleware framework, illustrating particular points within various policy and meta-policy dimensions. Section 4.2 shows how some of these Use Cases can be realised using our framework.

**UC1** A particular node hosts a number of services implemented by different objects. All data returned to Web Service clients should be transmitted *by value*, regardless of any conflicting policies specified by individual service implementations. Data returned to other types of clients should be transmitted *by reference*.

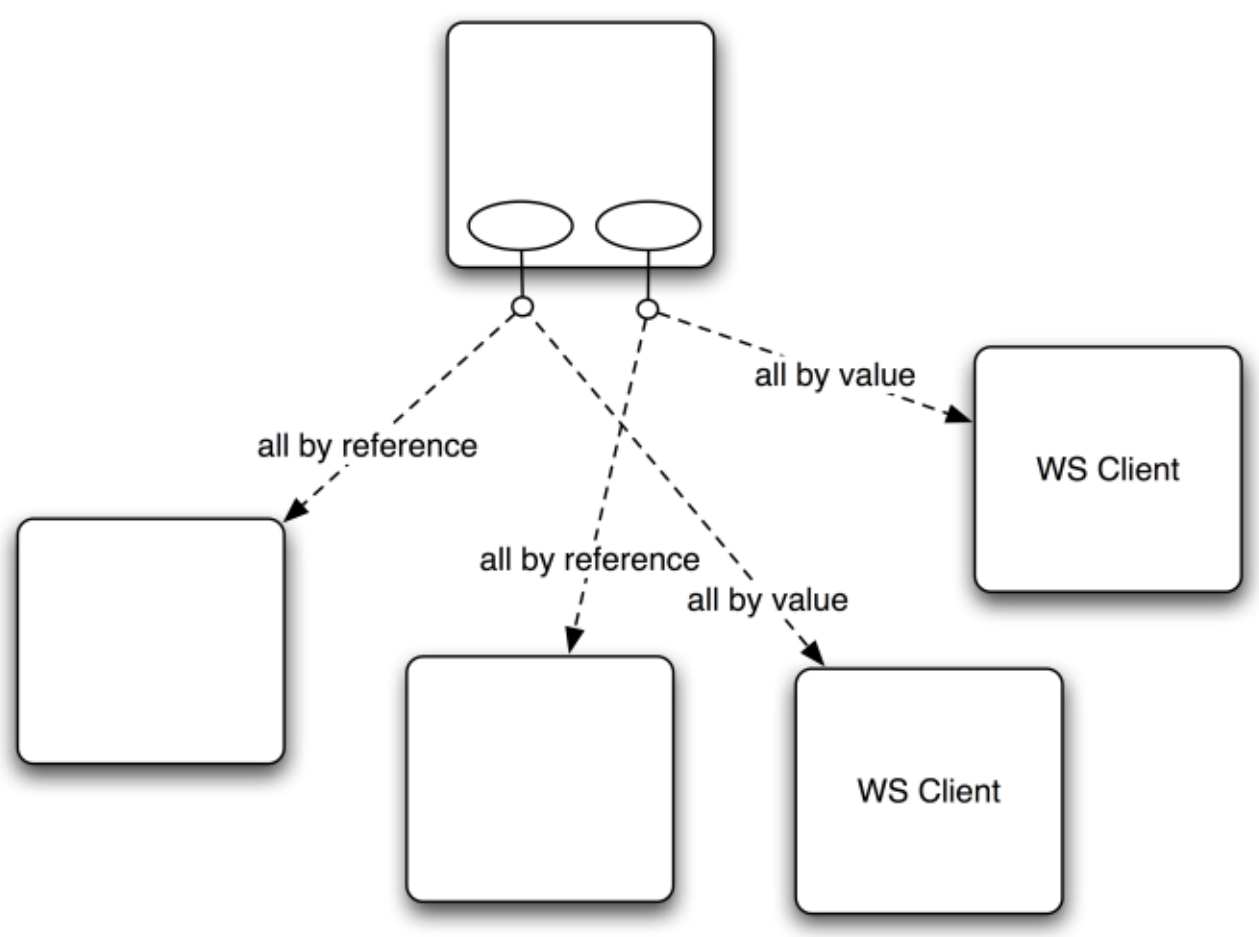

**Figure 2. UC1 - Varying policy by agent type.**

**UC1.1 [variant]** Data returned to a client instance known to have poor connectivity should be transmitted *by reference*, and *by value* to other client instances. Another potential motivation for this is to spread computational load, since an operation invoked on an object received *by reference* will be executed on the server, as opposed to locally on the client with *by value.*

**UC2** An object representing a node in a P2P system exposes two different services, *ViewNode* and *ManageNode*, with different types. Data returned by methods in the *ViewNode* service should be transmitted *by value*, so that the resulting copy of the data can be manipulated freely by the recipient without affecting the P2P node. Data returned by methods in the *ManageNode* service, access to which is subject to authentication, should be transmitted *by reference*, so that the original node can be controlled remotely.

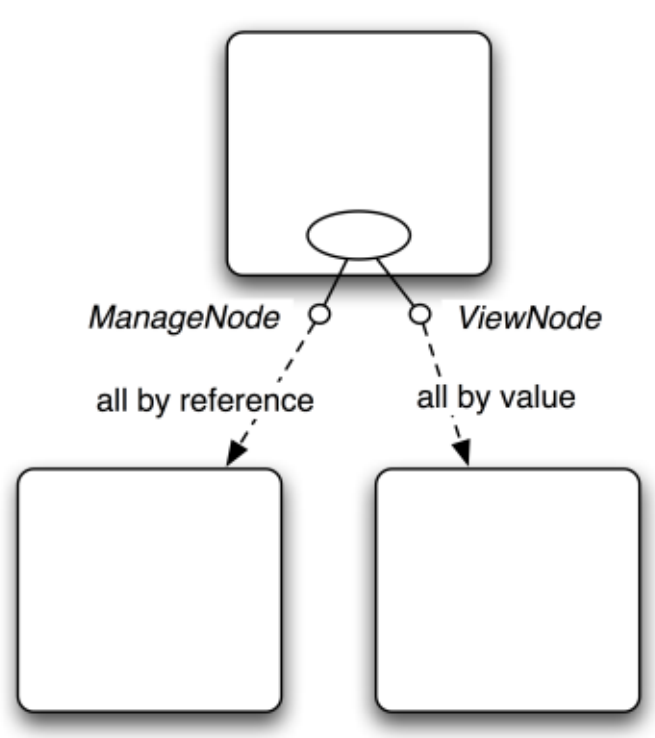

**Figure 3. Varying policy by service.**

**UC3** Instances of class *P2PNode*, representing nodes in a P2P system, should be transmitted *by reference* to ensure that each instance always resides on the host that it represents. To assist in error reporting, the value of the *key* field, containing the node's fixed P2P key, should be cached within the transmitted remote reference. This enables the holder of such a reference to access the key value even if the node or intervening network fails.

**UC4** All instances of class *Person* should be transmitted *by value*; data returned by methods in the *Directory* service should be transmitted *by reference*, except where this conflicts with the policy for class *Person*.

**UC4.1 [variant]** The policy for service *Directory* should take precedence over the policy for class *Person*.

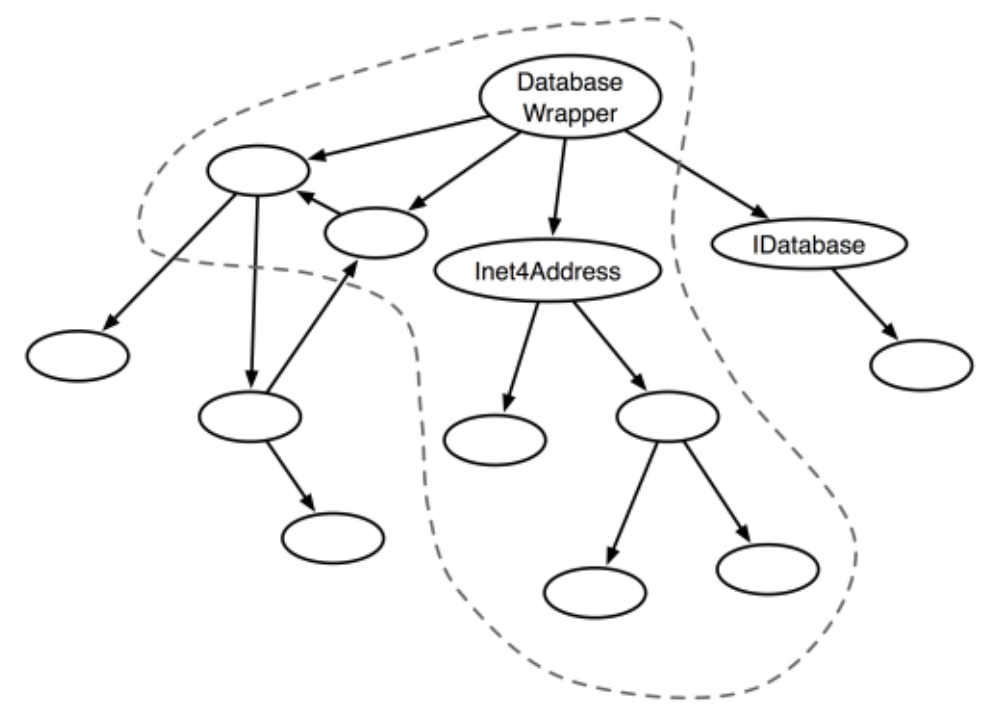

**Figure 4. Sub-graph to be transmitted by value.**

**UC5** Each instance of exact type *DatabaseWrapper* should be transmitted *by value*, together with the objects in its closure to a depth of 1. Each instance of type *InetAddress*, or a sub-type, should be transmitted *by value*, together with its entire closure. Each instance of interface type *IDatabase* should be transmitted *by reference*. The combination of these policies is shown in Figure 4, where the dotted line shows the sub-graph that should be transmitted *by value* with an instance of *DatabaseWrapper*.

**UC6** In order to constrain bandwidth consumption, each instance of type *JPEGImage* should be transmitted *by value* if the size of the image data is less than 500KB, and *by reference* otherwise.

**UC6.1 [variant]** Each instance of type *JPEGImage* should be re-encoded with a higher compression factor before transmission.

**UC7** Instances of class *Class* should always be transmitted *by value*. Rather than transmitting the state of a *Class* instance, it should be encoded simply as its fully qualified name, enabling the equivalent instance to be substituted on the receiving host.

**UC7.1 [variant]** Where an instance of class *Class* is transmitted to a host on which the corresponding class is not available, the encoding of the instance should include the bytes representing the class, enabling it to be dynamically loaded on the receiving host.

**UC8** By default, arrays that are transmitted *by value* are encoded in a SOAP-compatible format, using a separate XML element for each array element. This is a highly inefficient encoding for arrays with many small elements. Therefore, a byte array should be encoded as a single XML element containing the *base64* encoding of the entire data.

# 4. AN APPROACH TO PROVIDING POLICY-FREE ADAPTIVE MIDDLEWARE

Here we describe our particular approach to providing an adaptive policy framework. Section 2.5 listed some general requirements for such a framework. To these we add the following more specific requirements:

- policies must be *dynamically changeable*, meaning that the set of policies that are in place may be changed at any point before or during application execution;
- policies must be *capable of dynamic decision making*, implying that such policies must be supplied with appropriate information about the context in which computation occurs;
- a client application must be able to control the policies in effect on a given server that are pertinent to the client.

The last requirement arises since the policies in place in a given address-space affect only local operations. In the case of transmission and encoding policy, the local policies affect only outgoing transmissions, whether arguments to outgoing calls, or results of incoming calls. In order for a client to control the manner in which it receives results from calls to a server, it is necessary for it first to set the server's corresponding policy remotely.

## 4.1 Policy Management

Our adaptive policy framework meets the preceding requirements, and can succinctly specify all of the Use Cases described earlier. This is achieved using the *IMetaPolicyManager* interface shown in Figure 5, which provides methods for setting meta-policies in each of the policy dimensions identified in Section 2.3. For brevity we focus on the transmission policy dimension—the others are handled in a similar manner. The method *setTransmissionMetaPolicy* associates a particular transmission context with a corresponding policy.

```
public interface IMetaPolicyManager {
  void setTransmissionMetaPolicy(
    String contextPattern, TransmissionPolicy p,
    boolean matchSubtypes, TemporalScope s);
  void setEncodingMetaPolicy(
    String contextPattern, EncodingPolicy p,
    boolean matchSubtypes, TemporalScope s);
  void setPlacementMetaPolicy(
    String contextPattern, PlacementPolicy p,
    boolean matchSubtypes, TemporalScope s);
  ... // Other dimensions omitted for brevity.
  ... // Methods to iterate/remove rules omitted.
}
```

**Figure 5. Meta-policy management API.**

A simple pattern language[4] is used to specify contexts. The aim is that it should be easy to define simple policies, while being expressive enough to describe complex, static, dynamic, orthogonal and overlapping policies. The pattern language permits a context to be described in terms of any number of the meta-policy dimensions discussed in Section 2.4. A coarse-grain context, corresponding to a wide range of actual situations, is described by specifying details for only a small number of dimensions, while specifying many dimensions yields a highly specific fine-grain context. An example of a coarse-grain context pattern is:

```
object_type=a.b.C, method=d.e.F.m1()
```

This pattern would match contexts in which an instance of class *a.b.C* was being transmitted as a parameter to, or as the result from, the method *d.e.F.m1()*.

The full list of tags is: `thread`, `agent_type`, `agent_instance`, `parameter`, `method`, `service`, `field`, `object_type`, `root_type`, `package`. In addition to describing particular values for each of the meta-policy dimensions, the pattern language includes negation and two kinds of wildcard: '*' represents *always match*; '-' represents *default*. The former matches against the actual context in all cases, whereas the latter matches only if there is no other more specific pattern that would match. Examples are given in Section 4.2.

[4] Fully specified at http://… *[removed for anonymity]*.

The API with which particular transmission policies are specified is outlined in Figure 6. The base class, *TransmissionPolicy*, permits simple *by reference* and *by value* policies to be named. The class *ByValueToDepth* represents policies where full or partial object closures should be passed *by value*. *ByReferenceWithCaching* permits the specification of fields and methods to be cached on the client side when objects are transmitted *by reference*—this enables smart proxies such as those found in Orbix [9].

The final sub-class of *TransmissionPolicy* is *DynamicPolicy*, meeting the requirement for dynamic decision making. Unlike the others, this class is *abstract*, permitting application writers to define policies providing fine-grain adaptiveness. The *getPolicy* method takes as parameter an instance of *TransmissionContext* describing an actual transmission context. This supplies the implementation with sufficient information to be able to dynamically determine the appropriate policy based on, for example, caller, size of data or other application-specific requirements. It may be considered to be a reification [7] of the execution context for a particular object transmission.

The *setTransmissionMetaPolicy* method in Figure 5 takes another two parameters not yet discussed. The first is a flag specifying whether sub-type matching should be used when comparing the pattern against actual transmission contexts. For example, if *true* for a pattern specifying *Person* as the object type, then the pattern will match when sub-types of *Person* are transmitted. The final parameter specifies the temporal scope of the meta-policy rule. Various temporal scopes may be defined, the longest of which is *INDEFINITE*, which establishes the rule until explicitly removed, and the shortest of which is *CURRENT_CALL*, which establishes the rule until the inter-address-space call currently being serviced has returned. The latter permits adaptive behaviour to accommodate transient circumstances.

```
public class TransmissionPolicy {
  public static TransmissionPolicy BY_REF;
  public static TransmissionPolicy BY_VAL;
  protected TransmissionPolicy() {...}
}
public class ByValueToDepth
      extends TransmissionPolicy {
  public static ByValueToDepth FULL_CLOSURE;
  public int getDepth() {...}
  public ByValueToDepth(int depth) {...}
}
public class ByReferenceWithCaching
      extends TransmissionPolicy {
  public boolean isCached(Method method) {...}
  public boolean isCached(Field field) {...}
  public ByReferenceWithCaching(
    Field[] fields, Method[] methods) {...}
}
public abstract class DynamicPolicy
      extends TransmissionPolicy {
  public abstract TransmissionPolicy getPolicy(
    TransmissionContext c);
}
```

**Figure 6. Transmission policy API.**

The final requirement of the framework, that a client should be able to control the relevant policies on a server, can be met by making the meta-policy management interface for each address-space remotely accessible. Such management interfaces must themselves be subject to programmer-specifiable security policy.

## 4.2 Examples

In this section we show how selected Use Cases can be programmed using the adaptive policy framework[5]. All examples assume the prior declaration of *manager*, referring to the local instance of the meta-policy manager.

**UC1** The *by value* policy for Web Service clients is specified as[6]:

```
manager.setTransmissionMetaPolicy(
  "agent_type=WS, others=*",
  TransmissionPolicy.BY_VAL,
  false, TemporalScope.INDEFINITE);
```

The agent type identifier *WS* is matched against that specified by the remote client when it makes the call. The *others* tag is a short-hand for specifying all of the other context tags. Here the *always match* wildcard is used to ensure that this pattern will always be selected for Web Service clients, regardless of other contextual aspects such as the thread, the object type, etc. The value of the flag is unimportant since no type matching is involved here. The default *by reference* policy is specified as follows:

```
manager.setTransmissionMetaPolicy(
  "all=-", TransmissionPolicy.BY_REF,
  false, TemporalScope.INDEFINITE);
```

Here the *default* wildcard is specified for all of the contextual aspects, using the short-hand *all*, ensuring that the rule is selected only when no more specific rules are applicable.

**UC4** The following call sets a *by value* policy for all instances of class *Person* or any sub-class. As in *UC1*, the *always match* wildcards override all other contextual aspects.

```
manager.setTransmissionMetaPolicy(
  "object_type=Person, others=*",
  TransmissionPolicy.BY_VAL,
  true, TemporalScope.INDEFINITE);
```

A non-conflicting policy for the *Directory* service is specified as:

```
manager.setTransmissionMetaPolicy(
  "service=Directory, object_type=-, others=*",
  TransmissionPolicy.BY_REF,
  true, TemporalScope.INDEFINITE);
```

The *default* wildcard in the *object_type* element gives the rule lower precedence than other more specific rules. Thus the rule will be selected only in contexts where the service is *Directory* and the type of the object being transmitted is not *Person*.

This precedence order may be reversed, so that results from the *Directory* service are always transmitted *by reference* while *Person* instances are transmitted *by value* from all other services, by adjusting the wildcards as follows:

```
manager.setTransmissionMetaPolicy(
  "object_type=Person, service=-, others=*",
  TransmissionPolicy.BY_VAL,
  true, TemporalScope.INDEFINITE);
manager.setTransmissionPolicy(
  "service=Directory, others=*",
  TransmissionPolicy.BY_REF,
  true, TemporalScope.INDEFINITE);
```

It is also possible to set potentially conflicting policies:

```
manager.setTransmissionMetaPolicy(
  "object_type=Person, others=*",
  TransmissionPolicy.BY_VAL,
  true, TemporalScope.INDEFINITE);
manager.setTransmissionPolicy(
  "service=Directory, others=*",
  TransmissionPolicy.BY_REF,
  true, TemporalScope.INDEFINITE);
```

Here both patterns are intended to ensure precedence over other rules specifying different contextual aspects. A potential conflict arises when an instance of *Person* is returned by the *Directory* service. It is resolved using the following fixed ordering defined over contextual aspects, in order of decreasing precedence:

*thread, agent type, agent instance, parameter, method, service, field, object type, root object type, package*

Thus in this example, the rule specifying the service is selected over that specifying the object type.

**UC6** The dynamically evaluated policy necessary in this Use Case is specified as shown, assuming a method *getSize* to calculate the size of a given object. To create an instance of *DynamicPolicy* it is necessary to define the method *getPolicy*, which takes as parameter a description of the actual transmission context. In this example, the policy accesses the object that is about to be transmitted, via *context.currentObject*, calculates its size, and returns a *by value* or *by reference* policy as appropriate.

```
TransmissionPolicy bySize = new DynamicPolicy() {
  public TransmissionPolicy getPolicy(
      TransmissionContext context) {
    if (getSize(context.currentObject) < 500)
      return TransmissionPolicy.BY_VAL;
      else return TransmissionPolicy.BY_REF;
  }};
manager.setTransmissionMetaPolicy(
  "object_type=JPEGImage, others=*",
  bySize, false, TemporalScope.INDEFINITE);
```

# 5. RELATED WORK

Arguably, the first middleware system to support adaptiveness was CORBA [16]. Orbix [9] provided the concept of interceptors, which permitted the establishment of a chain of software components to handle outgoing requests. By default, the chain would typically hold a single interceptor that sent the request using the standard IIOP protocol, but several interceptors could be chained to add transaction information, encrypt the message, and send it using an arbitrary protocol. This pattern has been adopted by many systems since its invention. Whilst powerful, this scheme does not provide the degree of adaptability described here.

Our approach has much in common with AOP, in that it provides a mechanism for policy to be applied uniformly across an existing code base without the need to change that code. The main difference in emphasis is that we do not claim generality, instead focusing on dimensions of adaptiveness specific to middleware. Our framework is also more flexible than the common AOP platforms in that it allows policies to be changed dynamically.

---

[5] The full set of Use Cases is explained at http://... *[removed for anonymity]*.

[6] Such policies are commonly required by almost all applications; they are specified in configuration files and set by default. However, we believe that such policies should not be hard-wired into the framework and that application designers should have the freedom to specify and override them if required.

Sadjadi & McKinley present a taxonomy of adaptive middleware [18] which categorises middleware as ranging from *static* (adaptability applied at development time) through to *fully dynamic* (applied at run-time). Within these, adaptation ranges from *customizable* to *configurable*, and *tunable* to *mutable*, respectively. The work presented in this paper permits adaptation to be specified at either development time or at run-time, and permits policies to be dynamically defined and applied. As described in [18], many middleware adaptation mechanisms provide tunability via a two step process of static AOP at compile time and reflection at run-time. Such mechanisms are typified by DynamicTAO [12] which is capable of adapting its behaviour according to local policies. OpenORB [2] is an example of mutable middleware in which the middleware core can evolve. The aim of these systems is similar to ours, however, these systems do not provide the combination of being able to provide both blanket policies and fine grain control within a simple, efficient framework.

A number of middleware systems with varying degrees of adaptiveness regarding object transmission and placement policy have emerged. However, none support application adaptiveness to the degree proposed here. JavaParty [17] supports object placement policies that are associated with classes, limiting the flexibility of these policies. Pangaea [19] makes use of migration support in JavaParty, providing a policy mechanism that can determine when and to where object migration should take place. Migration policies cannot evolve at run-time or respond to application events. Using J-Orchestra [13], programmers can statically associate a *pass by move* policy with classes that support migration. FarGo [8] provides a Java RMI-based distributed object model that allows programmers to create classes with explicit support for migration. Classes that support remote access or migration must extend special interfaces and a special compiler is used to generate versions of these classes that are accessible remotely using Java RMI. FarGo allows types that represent different migration policies to be imposed onto references. By altering the types of references dynamically, programmers can define migration policies. Finally, JBoss [10] AOP Remoting uses aspect-oriented programming techniques to instrument instances of existing classes for remote access. Pass-by-value semantics are always employed and there is no control over object placement, remote instantiation or migration.

Using DCOM [3], programmers can instruct factories to instantiate components on remote machines by either identifying the machine explicitly or deferring to the *Service Control Manager (SCM)*. In this manner, object placement policies may be defined in terms of one-to-one mappings between component identifiers and machines. Like CORBA, DCOM determines parameter-passing semantics based on the statically defined IDL.

In the .Net framework, there are two conceptually different approaches to making instances of classes available: *web services* and *remoting*. Any suitably (statically) attributed class can be made available as a web service. The remoting infrastructure places semantic restrictions on the inheritance hierarchies of classes supporting remote access and tightly binds parameter-passing semantics to the distribution of the application. Programmers need not define separate interfaces for classes supporting remote access, though all remotely accessible classes must extend the special base class *MarshalByRefObject*.

# 6. IMPLEMENTATION

Here we briefly sketch our implementation approach. Whenever a new meta-policy rule is added using one of the methods in Figure 5, an entry is inserted into a data structure that represents the combination of all the extant rules. This data structure is consulted whenever a policy decision is required: it is traversed by the algorithm that matches actual context to the most appropriate meta-policy rule. Figure 7 shows the data structure after the following rules from UC1 and UC4 have been added:

```
agent_type=WS, others=*                    -> BY_VAL
all=-                                      -> BY_REF
object_type=Person, service=-, others=* -> BY_VAL
```

For simplicity, only five of the meta-policy dimensions are illustrated here. The context matching algorithm performs a depth-first traversal of the tree, searching for a path from the root to a leaf that matches all elements of the actual context. Each node in the tree contains up to four child entries, corresponding to the four types of pattern element: '*' (wildcard), specific value, '!' (negated specific value) and '-' (wildcard). Child entries are always traversed in this order, yielding the appropriate precedence. When the traversal reaches a node with no child entry matching the actual context, backtracking occurs. The traversal terminates when the first leaf is encountered, signifying that the appropriate policy has been located.

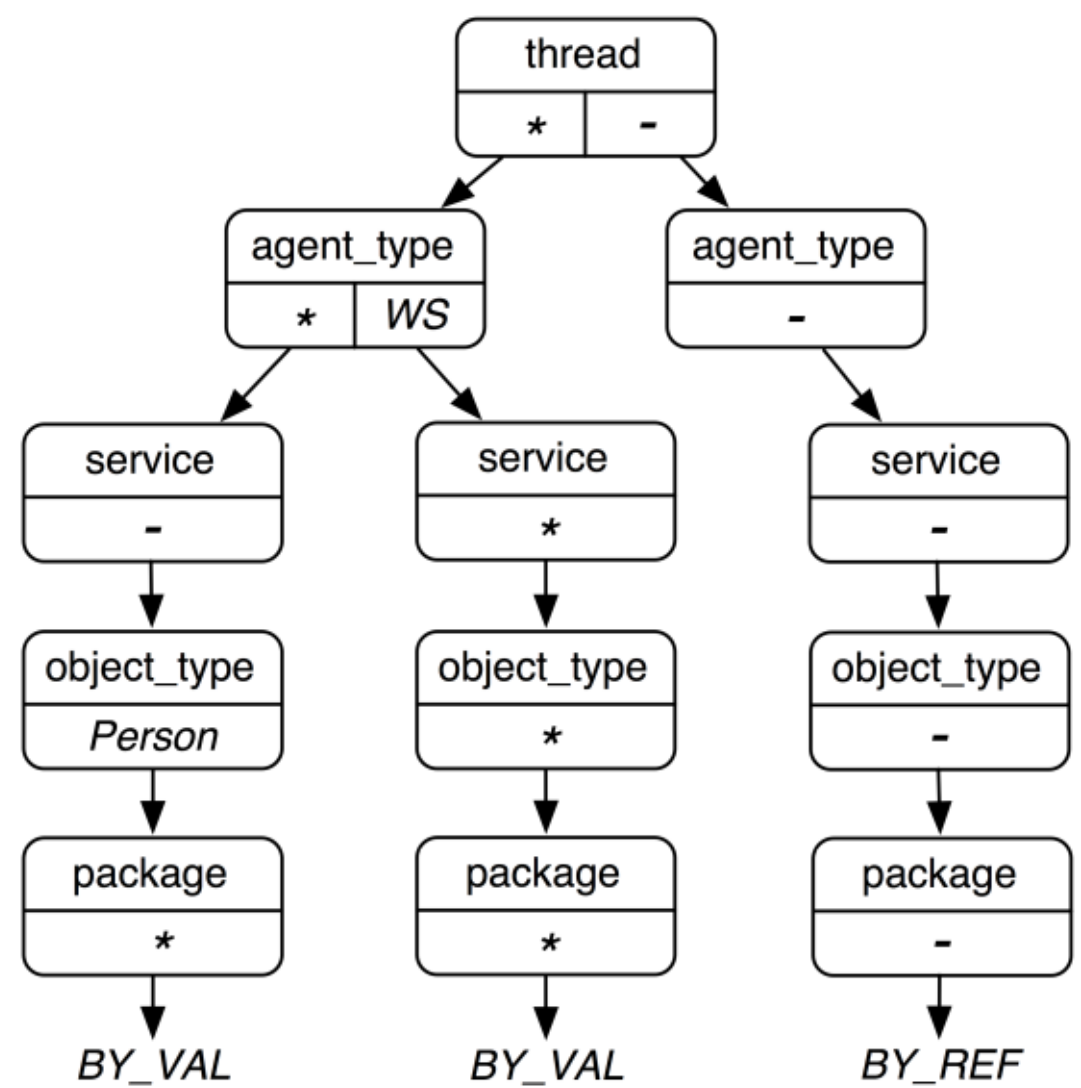

**Figure 7. Meta-policy data structure.**

For example, assume the following concrete transmission context:

```
thread identity = 318264
agent type = RAFDA
service = ManageNode
object type = HashMap
package = java.util
```

The traversal starts at the root node, matches * and moves to the root's first child where * is matched. In the *service* node the default is matched since it has no other children. On reaching the *object type* node there is no match since the types are incompatible and there are no wildcard entries. Backtracking leads first to the previously encountered *agent type* node, with no further matches, and then back to the root node. Here the default wildcard matches, and traversal proceeds all the way down the right-most branch of the tree to reach the appropriate policy, *BY_REF*.

An earlier and more restrictive version of this policy framework has been implemented in the context of RAFDA, a freely-available Java-based middleware system [5]. The framework described in this paper represents the latest iteration in our attempts to provide flexibility and adaptability. The RAFDA system has a number of uniquely combined properties crucial for supporting adaptive systems: it unifies the distributed object model and the service-oriented model; it supports adaptiveness with respect to the topology of distributed components and the manner in which components interact; finally it supports inter-operation with components written using industry-standard technologies. RAFDA is based on an HTTP/SOAP server architecture supporting adaptive encoding schemes; the system uses SOAP to interact with Web Service clients; arbitrary encoding schemes may be used with RAFDA-aware clients. RAFDA has been extensively used in distributed applications within various research projects.

## 7. CONCLUSIONS

This paper has motivated the need for adaptive policy-free middleware, and outlined an attempt to support this via a flexible policy framework. Our description has focused on object transmission and encoding policies, and the appropriate meta-policy dimensions for specifying how policies should be automatically selected according to dynamic context. We believe that this approach is also suitable for wider application in other policy dimensions; we are currently investigating possibilities within the RAFDA middleware system.